\documentclass[aps,prl,twocolumn,superscriptaddress]{revtex4}
\usepackage{graphicx}
\usepackage{amsmath, amsthm, amssymb}
\usepackage{mathrsfs}
\usepackage{caption}
\usepackage[justification=raggedright,singlelinecheck=false]{caption}
\newcommand{\bra}[1]{\langle#1|}
\newcommand{\ket}[1]{|#1\rangle}
\DeclareMathSizes{10}{10}{5}{3}

\begin{document} 
\title{A 2D Quantum Walk Simulation of Two-Particle Dynamics} 

\author{Andreas Schreiber}
	\email{Andreas.Schreiber@upb.de}
\affiliation{University of Paderborn, Applied Physics,Warburger Stra\ss e 100, 33098 Paderborn, Germany}
\affiliation{Max Planck Institute for the Science of Light, G{\"u}nther-Scharowsky-Stra\ss e 1/Bau 24,91058 Erlangen, Germany}
\author{ Aur{\' e}l G{\' a}bris}
\affiliation{Department of Physics, Faculty of Nuclear Sciences and Physical Engineering,Czech Technical University in Prague, B{\v r}ehov{\' a} 7, 115 19 Praha, Czech Republic} 
\affiliation{Wigner Research Centre for Physics, Hungarian Academy of Sciences,  H-1525 Budapest, P.O.Box 49, Hungary}
\author{Peter P. Rohde}
\affiliation{Centre for Engineered Quantum Systems, Department of Physics and Astronomy, Macquarie University, Sydney NSW 2113, Australia}
\author{Kaisa Laiho}
\affiliation{University of Paderborn, Applied Physics,Warburger Stra\ss e 100, 33098 Paderborn, Germany}
\affiliation{Max Planck Institute for the Science of Light, G{\"u}nther-Scharowsky-Stra\ss e 1/Bau 24,91058 Erlangen, Germany}
\author{Martin {\v S}tefa{\v n}{\' a}k}
\affiliation{Department of Physics, Faculty of Nuclear Sciences and Physical Engineering,Czech Technical University in Prague, B{\v r}ehov{\' a} 7, 115 19 Praha, Czech Republic} 
\author{V{\'a}clav Poto{\v c}ek}
\affiliation{Department of Physics, Faculty of Nuclear Sciences and Physical Engineering,Czech Technical University in Prague, B{\v r}ehov{\' a} 7, 115 19 Praha, Czech Republic} 
\author{Craig Hamilton}
\affiliation{Department of Physics, Faculty of Nuclear Sciences and Physical Engineering,Czech Technical University in Prague, B{\v r}ehov{\' a} 7, 115 19 Praha, Czech Republic} 
\author{ Igor Jex}
\affiliation{Department of Physics, Faculty of Nuclear Sciences and Physical Engineering,Czech Technical University in Prague, B{\v r}ehov{\' a} 7, 115 19 Praha, Czech Republic} 
\author{Christine Silberhorn }
\affiliation{University of Paderborn, Applied Physics,Warburger Stra\ss e 100, 33098 Paderborn, Germany}
\affiliation{Max Planck Institute for the Science of Light, G{\"u}nther-Scharowsky-Stra\ss e 1/Bau 24,91058 Erlangen, Germany}

\begin{abstract}
Multi-dimensional quantum walks can exhibit highly non-trivial topological structure, providing a powerful tool for simulating quantum information and transport systems. We present a flexible implementation of a 2D optical quantum walk on a lattice, demonstrating a scalable quantum walk on a non-trivial graph structure. We realized a coherent quantum walk over 12 steps and 169 positions using an optical fiber network. With our broad spectrum of quantum coins we were able to simulate the creation of entanglement in bipartite systems with conditioned interactions. Introducing dynamic control allowed for the investigation of effects such as strong non-linearities or two-particle scattering. Our results illustrate the potential of quantum walks as a route for simulating and understanding complex quantum systems.
\end{abstract}
\maketitle 

Quantum simulation constitutes a paradigm for developing our understanding of quantum mechanical systems. A current challenge is to find schemes, that can be readily implemented in the laboratory to provide insights into complex quantum phenomena. Quantum walks \cite{Aharonov93,Aharonov01,Kempe2003} serve as an ideal test-bed for studying the dynamics of such systems. Examples include understanding the role of entanglement and interactions between quantum particles, the occurrence of localization effects \cite{Konno2004}, topological phases \cite{Kitagawa2010}, energy transport in photosynthesis \cite{Mohseni2008,Plenio2008}, and  the mimicking of the formation of molecule states \cite{Werner11}. While theoretical investigations already take advantage of complex graph structures in higher dimensions, experimental implementations are still limited by the required physical resources.

All demonstrated quantum walks have so far been restricted to evolution in one dimension. They have been realized in a variety of architectures, including photonic \cite{bouwmeester1999,Schreiber10,Broome10,Schreiber11} and atomic \cite{schaetz09,karski09,zaehringer10} systems. Achieving increased dimensionality in a quantum walk \cite{Sanders} is of practical interest as many physical phenomena cannot be simulated with a single walker in a one-dimensional quantum walk, such as multi-particle entanglement and non-linear interactions. Furthermore, in quantum computation based on quantum walks \cite{Childs09, Lovett2010} search algorithms exhibit a speed-up only in higher dimensional graphs \cite{childs2003,SKW,ambainis2004,Hillery2010}. The first optical approaches to increasing the complexity of a linear quantum walk \cite{Peruzzo10,OBrien2011,Osellame_2011} showed that the dimensionality of the system is effectively expanded by using two walkers, keeping the graph one-dimensional. While adding additional walkers to the system is promising, introducing conditioned interactions and in particular controlled non-linear interactions at the single photon level is technologically very challenging. Interactions between walkers typically result in the appearance of entanglement, and have been shown to improve certain applications, such as the graph isomorphism problem \cite{Wang11}. In the absence of such interactions, the two walkers remain effectively independent, which severely limits observable quantum features.

We present a highly scalable implementation of an optical quantum walk on two spatial dimensions for quantum simulation, using frugal physical resources. One major advance of a two-dimensional system is the possibility to simulate a discrete evolution of two-particles including controlled interactions. In particular, one walker, in our case a coherent light pulse, on a 2D lattice is topologically equivalent to two-walkers acting on a one-dimensional graph. Thus, despite using an entirely classical light source, our experiment is able to demonstrate several archetypal two-particle quantum features. For our simulations we exploit the similarity between coherent processes in quantum mechanics and classical optics \cite{Spreeuw,Kwiat1998}, as it was used for example to demonstrate Grover's quantum search algorithm \cite{Spreeuw2002}.

\begin{figure*}
\begin{center} 
\includegraphics[width=\textwidth]{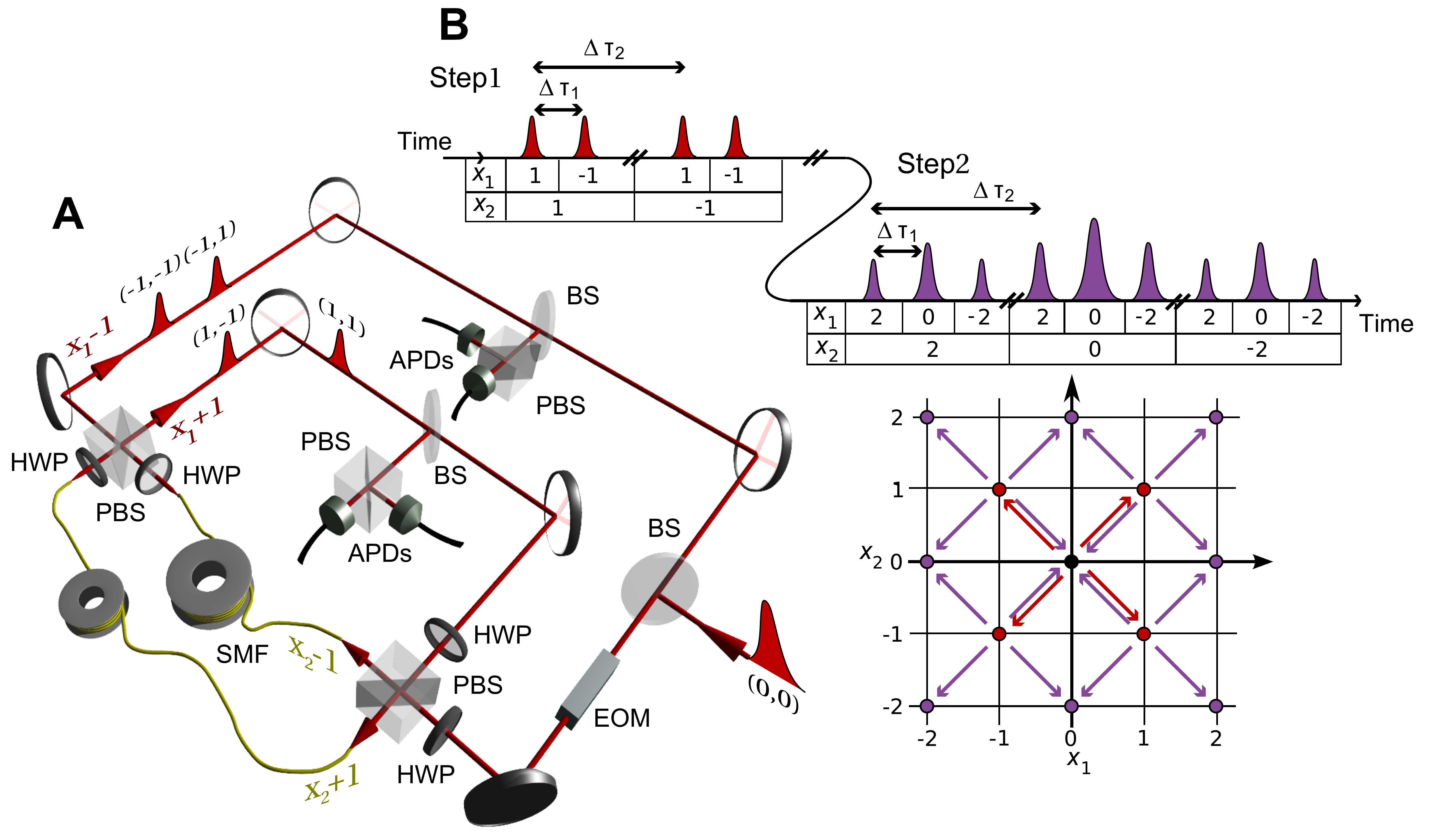}
\end{center}
\caption{ (\textbf{A}) Experimental setup. Our photon source is a pulsed diode laser with pulse width 88ps, wavelength 805nm and repetition rate 110kHz. The photons are initialized at position $\ket{x_1,x_2} = \ket{0,0}$ in horizontal polarization (corresponding to coin state $ \ket{c_1,c_2}=\ket{-1,-1}$). Once coupled into the setup through a low reflectivity beam splitter (BS, reflectivity $3\%$), their polarization state is manipulated with an EOM and a half-wave plate (HWP). The photonic wave packets are split by a polarizing beam splitter (PBS) and routed through single-mode fibres (SMF) of length 135m or 145m, implementing a temporal step in the $x_2$ direction. Additional HWPs and a second PBS perform a step in the $x_1$ direction based on the same principle. The split wave packet after the first step with equal splitting is indicated in the picture. At each step the photons have a probability of $12\%$ $(4\%)$ in loops $x_1-1$ ($x_1+1$) of being coupled out to a polarization and hence coin state resolving detection of the arrival time via four avalanche photodiodes (APDs). Including losses and detection efficiency, the probability of a photon continuing the walk after one step is $52\%$ $(12\%)$ without (with) the EOM. (\textbf{B}) Projection of the spatial lattice onto a one-dimensional temporally encoded pulse chain for step one and two. Each step consists of a shift in both $x_1$ direction, corresponding to a time difference of $\Delta \tau_1= 3.11$ns, and $x_2$ direction with $\Delta \tau_2 =46.42$ns.} 
\end{figure*}

A quantum walk consists of a walker, such as a photon or an atom, which coherently propagates between discrete vertices on a graph. A walker is defined as a bipartite system consisting of position ($x$) and a quantum coin ($c$). The position value indicates at which vertex in the graph the walker resides, while the coin is an ancillary quantum state determining the direction of the walker at the next step. In a two-dimensional quantum walk the basis states of a walker are of the form $\ket{x_1,x_2,c_1,c_2}$  describing its position $x_{1,2}$ in spatial dimension 1 and 2 and the corresponding two-sided coin parameters with $c_{1,2} =\pm1$. The evolution takes place in discrete steps, each of which has two stages, defined by coin ($\hat{C}$) and step ($\hat{S}$) operators. The coin operator coherently manipulates the coin parameter, leaving the position unchanged, whereas the step operator updates the position according to the new coin value. Explicitly, with a so-called Hadamard coin $\hat{C_H}=\hat{H}_{1}\otimes \hat{H}_{2}$, a single step in the evolution is defined by the operators,
\begin{eqnarray}
\hat{H_i}\ket{x_i,\pm 1} &\to& (\ket{x_i,1} \pm \ket{x_i,-1})/\sqrt{2}, \;\;\; \forall i= 1,2\nonumber \\
\hat{S}\ket{x_1,x_2,c_1,c_2} &\to& \ket{x_1+c_1,x_2+c_2,c_1,c_2}.
\end{eqnarray}
\label{eq}
The evolution of the system proceeds by repeatedly applying coin and step operators on the initial state $\ket{\psi_\mathrm{in}}$, resulting in $\ket{\psi_n}=(\hat{S}\hat{C})^n\ket{\psi_\mathrm{in}}$ after $n$ steps. The step operator $\hat{S}$ hereby translates superpositions and entanglement between the coin parameters directly to the spatial domain, imprinting signatures of quantum effects in the final probability distribution.

We performed 2D quantum walks with photons obtained from attenuated laser pulses. The two internal coin states are represented by two polarization modes (horizontal and vertical) in two different spatial modes [App. 1], similar to the proposal in \cite{Roldan2005}. Incident photons follow, depending on their polarization, four different paths in a fiber network (Fig. 1A). The four paths correspond to the four different directions a walker can take in one step on a 2D lattice. Different path lengths in the circuit generate a temporally encoded state, where different position states are represented by discrete time-bins (Fig. 1B). Each round trip in the setup implements a single step operation while the quantum coin operation is performed with linear optical elements (half-wave plates, HWP) [App. 1]. In order to adjust the coin operator independently at each position we employed a fast-switching electro-optic modulator (EOM). A measurement with time-resolving single-photon counting modules allowed for the reconstruction of the output photo-statistics [App. 2].

We have implemented two different kinds of quantum coins in our 2D quantum walks. First we investigated quantum walks driven only by separable coin operations, $\hat{C} = \hat{C}_1 \otimes \hat{C}_2$. Here the separability can directly be observed in the spatial spread over the lattice, when initializing the walker in a separable state. As an example we measured a Hadamard walk with photons initially localized at position $\ket{x_1,x_2}=\ket{0,0}$. The probability distribution showing at which position the photons were detected after ten steps (Fig. 2A+B) can be factorized into two independent distributions of one-dimensional quantum walks \cite{Sanders}, stating no conceptual advantage of a 2D-quantum walk.
However, two-dimensional quantum walks allow for much greater complexity using controlled operations. These operations condition the transformation of one coin state on the actual state of the other. Due to the induced quantum correlations one obtains a non-trivial evolution resulting in an inseparable final state. The probability distribution for a Hadamard walk with an additional controlling operation can be seen in Fig. 2C+D.
We compare the ideal theoretical distribution with the measured photo-statistics via the similarity, $S = (\sum\limits_{x_1,x_2} \sqrt{P_\mathrm{th}(x_1,x_2)P_\mathrm{exp}(x_1,x_2)})^2$, quantifying the equality of two classical probability distributions ($S=0$ for a completely orthogonal distributions and $S=1$ for identical distributions). For the Hadamard walk (Fig. 2A+B) we observe $S= 0.957 \pm 0.003$, and for the quantum walk with controlling gates (Fig. 2C+D) $S= 0.903 \pm 0.018 $ (after 10 steps, across 121 positions).

\begin{figure*}
\centering 
\includegraphics[width=0.9\textwidth]{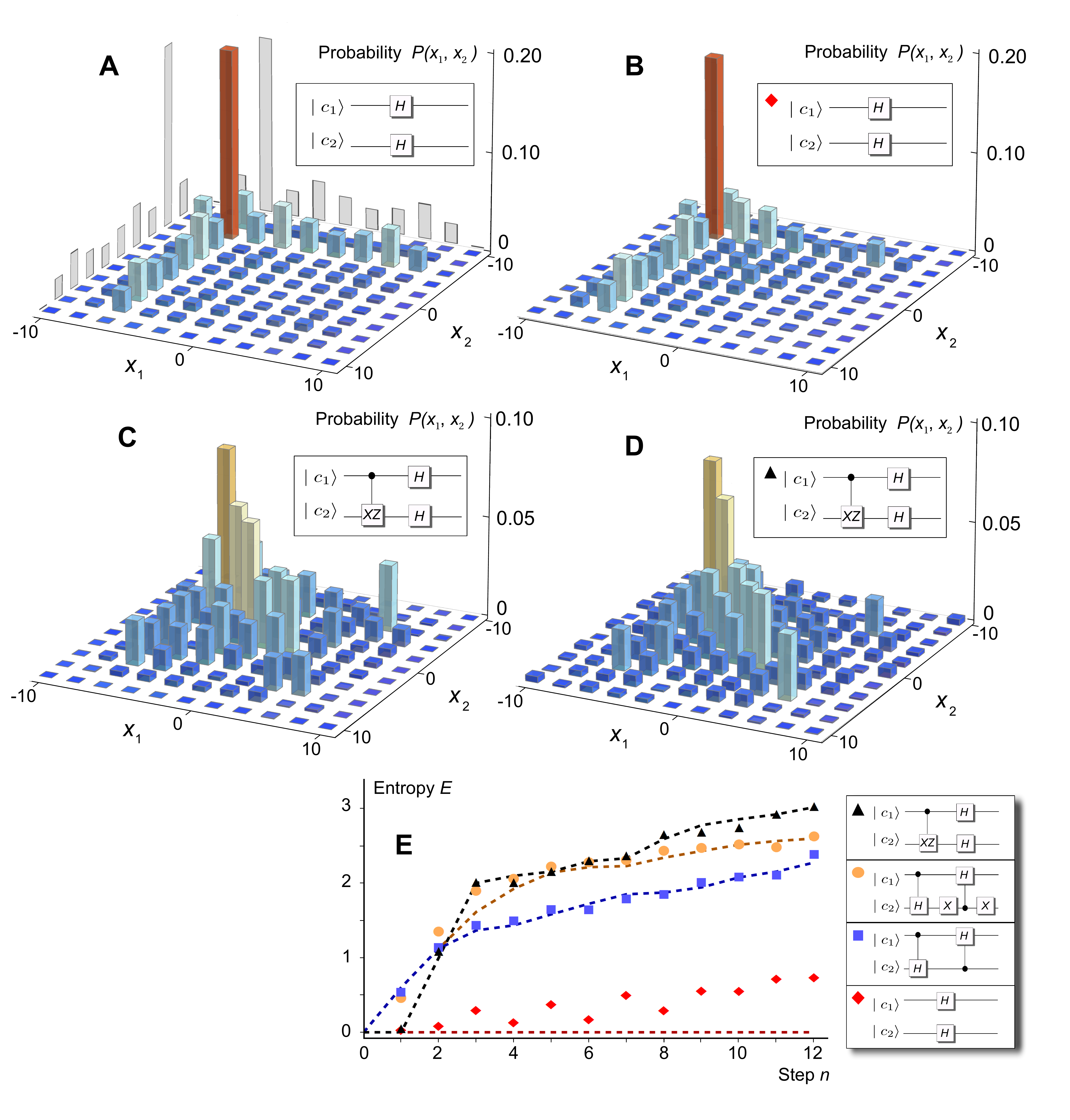}
\caption{ Measured and simulated probability distribution $P(x_1,x_2)$ (traced over the coin space) after ten steps of a 2D quantum walk with initial state $\ket{0,0,-1,-1}$. Theoretical (\textbf{A}) and measured (\textbf{B}) probability distribution of a 2D Hadamard walk using the operation $\hat{C}_H$ (Eq. 1). As only separable coin operations were performed (inset), the distribution is separable, given by a product of two one-dimensional distributions (gray). Theoretical (\textbf{C}) and measured (\textbf{D}) probability distribution of a 2D walk with controlled-Not $X$ and controlled-phase operation $Z$, resulting in an unfactorizable distribution. Here $c_2$ is only transformed by $XZ \ket{\pm1}\rightarrow \pm\ket{\mp1}$ if $c_1 = -1$ . The results (\textbf{B}) and (\textbf{D}) are obtained by detecting over $7\times 10^3$ events and calibrated by the detection efficiencies of all four coin basis states. (\textbf{E}) Dynamic evolution of the Von Neumann entropy E generated by quantum walks (\textbf{B}) and (\textbf{D}) and quantum walks using controlled Hadamard coin operations (inset). The experimental values (dots) and theoretical predictions (dashed lines) mark a lower boundary for simulated two-particle entanglement. Statistical errors are smaller than the dot size.} 
\end{figure*}

Increasing the number of walkers in a quantum walk effectively increases its dimensionality \cite{Peruzzo10}. Specifically, for a given 1D quantum walk with $N$ positions and two walkers, there exists an isomorphic square lattice walk of size $N^2$ with one walker. By this topological analogy, a measured spatial distribution from a 2D lattice with positions ($x_1,x_2$) can be interpreted as a coincidence measurement for two walkers at positions $x_1$ and $x_2$ propagating on the same linear graph. Hereby each combined coin operation of both particles, including controlled operations, has an equivalent coin operation in a 2D quantum walk. This allows us to interpret the 2D walk in Fig. 2C+D as a quantum walk with controlled two-particle operations, a system typically creating two-particle entanglement. The inseparability of the final probability distribution is then a direct signature of the simulated entanglement.

In Fig. 2E we show a lower bound for the simulated entanglement between the two particles during the stepwise evolution with four different coin operations. We quantified the simulated entanglement via the von Neumann entropy $E$, assuming pure final states after the quantum walk [App. 4]. For this calculation the relative phases between the positions and coins were reconstructed from the obtained interference patterns, while phases between the four coin states were chosen to minimize the entanglement value. Without conditioned operations the two particles evolve independently ($E=0$), whereas an evolution including controlled operations reveals a probability distribution characterized by bipartite entanglement. We found that the interactions presented in Fig. 2C+D exhibit an entropy of at least $E= 2.63\pm 0.01$ after 12 steps, which is $56\%$ of the maximal entropy (given by a maximally entangled state). The non-zero entropies obtained in the higher steps of the separable Hadamard walk are attributed to the high sensitivity of the entropy measure to small errors in the distribution for $E \approx 0$. 

The investigated interactions can be interpreted as long-distance interactions with the interaction strength being independent of the spatial distance of the particles. This is a unique effect and highly non-trivial to demonstrate in actual two-particle quantum systems.

Contrary to the position independent interactions is the evolution of two-particle quantum walks with short-range interactions, that is interactions occurring only when both particles occupy the same position. These interactions can be interpreted as two-particle scattering or non-linear interactions. When utilizing a 2D quantum walk to simulate two walkers, all vertices on the diagonal of the 2D-lattice correspond to both walkers occupying the same position. Hence, we can introduce non-linear interactions by modifying the coin operator on the diagonal positions while keeping all other positions unaffected. As an example of a two-particle quantum walk with non-linear interactions (Fig. 3), the coin operator on the diagonal is in the form $C_{\textrm{nl}}= (H_1 \otimes H_2) C_{Z}$, where $C_{Z}$ is a controlled phase operation implemented by a fast switching EOM. The chosen operation simulates a quantum scenario of particular interest -- the creation of bound molecule states, predicted as a consequence of two-particle scattering \cite{Werner11}. Evidently, the quantum walk is to a large extent confined to the main diagonal ($\sum\limits_x P(x,x)=0.317\pm0.006$ as opposed to the Hadamard walk $\sum\limits_x P(x,x)= 0.242\pm0.001$), a signature of the presence of a bound molecule state. In general, using a coin invariant under particle exchange, bosonic or fermionic behavior can be simulated, depending on whether the initial states are chosen to be symmetric or anti-symmetric with respect to particle permutations. With our initial state being invariant under particle exchange we simulated an effective Bose-Hubbard type non-linearity for two bosons \cite{Lahini2011}. 
\begin{figure*}
\centering 
\includegraphics[width=0.8\textwidth]{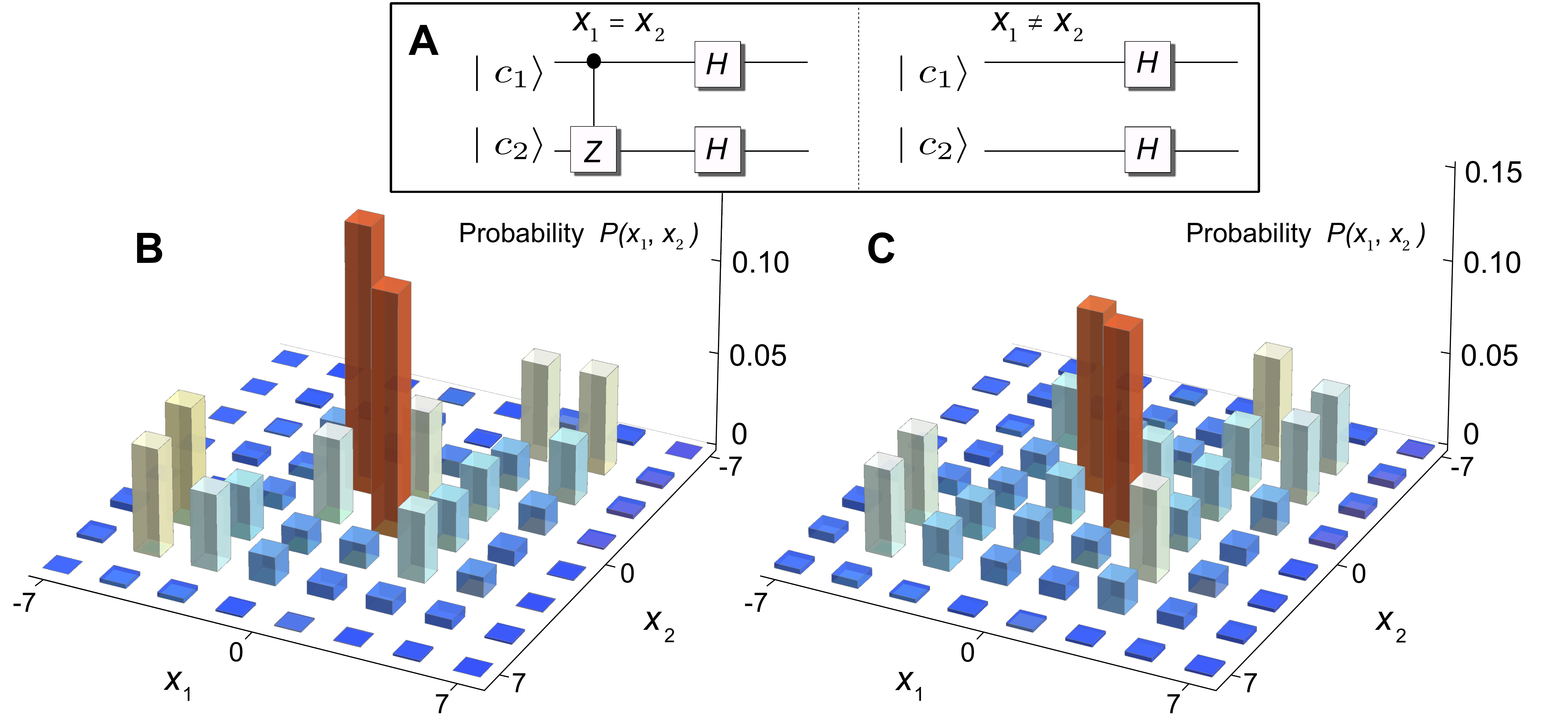}
\caption{ (\textbf{A}) Circuit representation of coin operations simulating non-linear interactions via 2D quantum walk. Only when the two virtual particles meet ($x_1=x_2$) a controlled operation is applied. Theoretical (\textbf{B}) and measured (\textbf{C}) coincidence distribution $P(x_1,x_2)$ (traced over the coin space) after seven steps of a simulated two-particle quantum walk with initial state $\ket{0,0,-1,-1}$. The high probability that both particles are at the same position (diagonal) is a striking signature of bound states.
The measured distribution is reconstructed by detecting over $8\times 10^3$ events and has a similarity of $S=0.957\pm0.013 $. Adding the EOM to the setup for dynamical control limits the step number to $n=7$ due to the higher losses per step. Small imperfections of the EOM are included in the theoretical plot. } 
\end{figure*}
We have demonstrated an efficient implementation of a two-dimensional quantum walk and proved the experimental feasibility to simulate a diversity of interesting multi-particle quantum effects. Our experiment overcomes the technical challenges of two-particle experiments, while exhibiting very high similarity and scalability. Combined with the flexibility in the choice of input state, controlling the coin at each position independently allows for simulations of a broad spectrum of dynamic quantum systems under different physical conditions. 

Our experimental architecture can be generalized to more than two dimensions, with the addition of extra loops and orbital angular momentum modes as coin states \cite{hamilton11}. This opens a largely unexplored field of research, facilitating quantum simulation applications with multiple walkers, including bosonic and fermionic behavior, and non-linear interactions. It may be possible to study the effects of higher dimensional localization, graph percolations or utilize the network topology in conjunction with single- or two-photon states. Additionally, a foreseeable future application for our system is the implementation of a quantum search algorithm. We demonstrated that, with a physical resource overhead, a classical experiment can simulate many genuine quantum features. While our experiment is important for simulation applications, it is equally interesting for understanding fundamental physics at the border between classical and quantum coherence theory.

\newpage
\noindent\textbf{Acknowledgements:} We acknowledge financial support from the German Israel Foundation (Project No. 970/2007). AG, M{\v S}, VP, CH and IJ acknowledge grant support from MSM6840770039 and MSMT LC06002, SGS10/294/OHK4/3T/14, GA \v CR 202/08/H078 and OTKA T83858. PR acknowledges support from the Australian Research Council Centre of Excellence for Engineered Quantum Systems (Project number CE110001013).
\clearpage
\begin{appendix}
\renewcommand{\thefigure}{A.\arabic{figure}}
\setcounter{figure}{0} 
\section{Appendix 1: Quantum gates with optical elements}\label{App. 1}
We realized the four internal coin states $\ket{\pm1,\pm1}$ with the linear polarization states, horizontal $\ket{H}$ and vertical $\ket{V}$, and two spatial modes $\ket{a}$ and $\ket{b}$, similar to the four spatial modes proposed in (\textit{27}). The spatial modes correspond to the two input ports of the first polarizing beam splitter (Fig.~A.1)). We encoded the states by 
\begin{align*}
\ket{H,a}\rightarrow\ket{-1,-1}; \ \ket{V,a}\rightarrow\ket{-1,+1};\\
\ \ket{H,b}\rightarrow\ket{+1,+1};\ \ket{V,b}\rightarrow\ket{+1,-1}.
\tag{A1}
\end{align*}
To implement our quantum operations in the four-dimensional Hilbert space of the quantum coin we decomposed the $U(4)$ unitary coin operation into products of multiple $U(2)$ operations (\textit{31}).
Each $U(2)$ transformation is implemented either by half-wave plates (HWP) or an electro-optic modulator (EOM).

In the basis of the four coin states the transformations are given by
\begin{equation*}
\begin{array}{l}
\hat{C}_{\textrm{HWP}_1} = {\left(\begin{array}{cccc}%
                           \cos(2 \theta_1) & \sin(2 \theta_1)&0&0  \\
                           \sin(2 \theta_1) & -\cos(2 \theta_1)&0&0 \\
				0&0&1&0\\
				0&0&0&1
                 \end{array}\right)}, \\\\
 \hat{C}_{\textrm{HWP}_2} = {\left(%
                 \begin{array}{cccc}
      			1&0&0&0\\
			0&1&0&0\\
			0&0&  \cos(2 \theta_2) & \sin(2 \theta_2) \\
                     0&0& \sin(2 \theta_2) & -\cos(2 \theta_2)
                 \end{array}%
               	 \right)},\\\\
\hat{C}_{\textrm{HWP}_3} = {\left(%
                 \begin{array}{cccc}
                           \cos(2 \theta_3) & 0&0&\sin(2 \theta_3)  \\
                            0&1&0&0\\
				0&0&1&0\\
				 \sin(2 \theta_3) &0& 0 & -\cos(2 \theta_3)
                 \end{array}%
               	 \right)},\\\\
\hat{C}_{\textrm{HWP}_4} =  {\left(%
                 \begin{array}{cccc}
                      1 &0&0&0  \\
                      0&-\cos(2 \theta_4) & \sin(2 \theta_4)&0 \\
			0&\sin(2 \theta_4) & \cos(2 \theta_4)&0\\
			0&0&0&1
                 \end{array}%
               	 \right)},\\\\
\hat{C}_{\textrm{EOM}} =   {\left(%
                 \begin{array}{cccc}
                           e^{\imath \phi(x_1,x_2)} & 0 &0&0 \\
                           0 & 1&0&0 \\
				0&0&1&0\\
				0&0&0&1
                 \end{array}%
               	 \right)},

\end{array}
\tag{A2}
\end{equation*}
with $\theta_i$ being the angle of $\textrm{HWP}_i$, $i=\{1,..,4\}$, relative to its optical axis and $\phi(x_1,x_2)$ a tunable phase.

Depending on the position of the four HWPs in the setup (Fig.~A.1A) they operate on different coin state pairs, due to the spatial switch via polarizing beam splitters (PBS).
The Hadamard coin (Fig.~2B+E) was obtained for the configurations $\theta_i =\pi/8,  \forall i$, while the coin used in Fig.~2D+E was given by $\theta_1 =-\pi/8$, $\theta_{2-4} =\pi/8 $. To implement controlled-Hadamard gates either $\textrm{HWP}_{2,3}$ or $\textrm{HWP}_{2,4}$  were aligned to their optical axis ($\theta=0$), while the remaining plates were set to $\theta =\pi/8$ (Fig.~2E). 
The transformation of the EOM with $\phi(x_1,x_2)= \pi$  corresponds to a controlled-Z operation.

Additional static phase factors changing the relative phase between the four coin states can occur during the propagation through the setup. However, these phases do not influence the final propability distribution due to the property of the coin operators that can be implemented with the used optical elements. Given precise phase control, two additional HWPs and a PBS would allow the implementation of arbitrary $U(4)$ coin operators.

\section{Appendix 2: Quantum walk implementation via time-multiplexing}\label{App. 2}

Our experiment simulates a 2D quantum walk on a regular square lattice, which means that a walker can move in four possible directions from a given site. The direction of the movement is determined by the current coin state of the walker. To implement the quantum walk in a 2D topology we use the time-multiplexing technique (\textit{9}). This method maps each individual position of the 2D graph on the one-dimensional time line. In contrast to determining the direction of the following step in space, the coin state defines a fixed time delay in the time-multiplexed system. 

\begin{figure*}
\begin{center}
\includegraphics[width =0.73\textwidth]{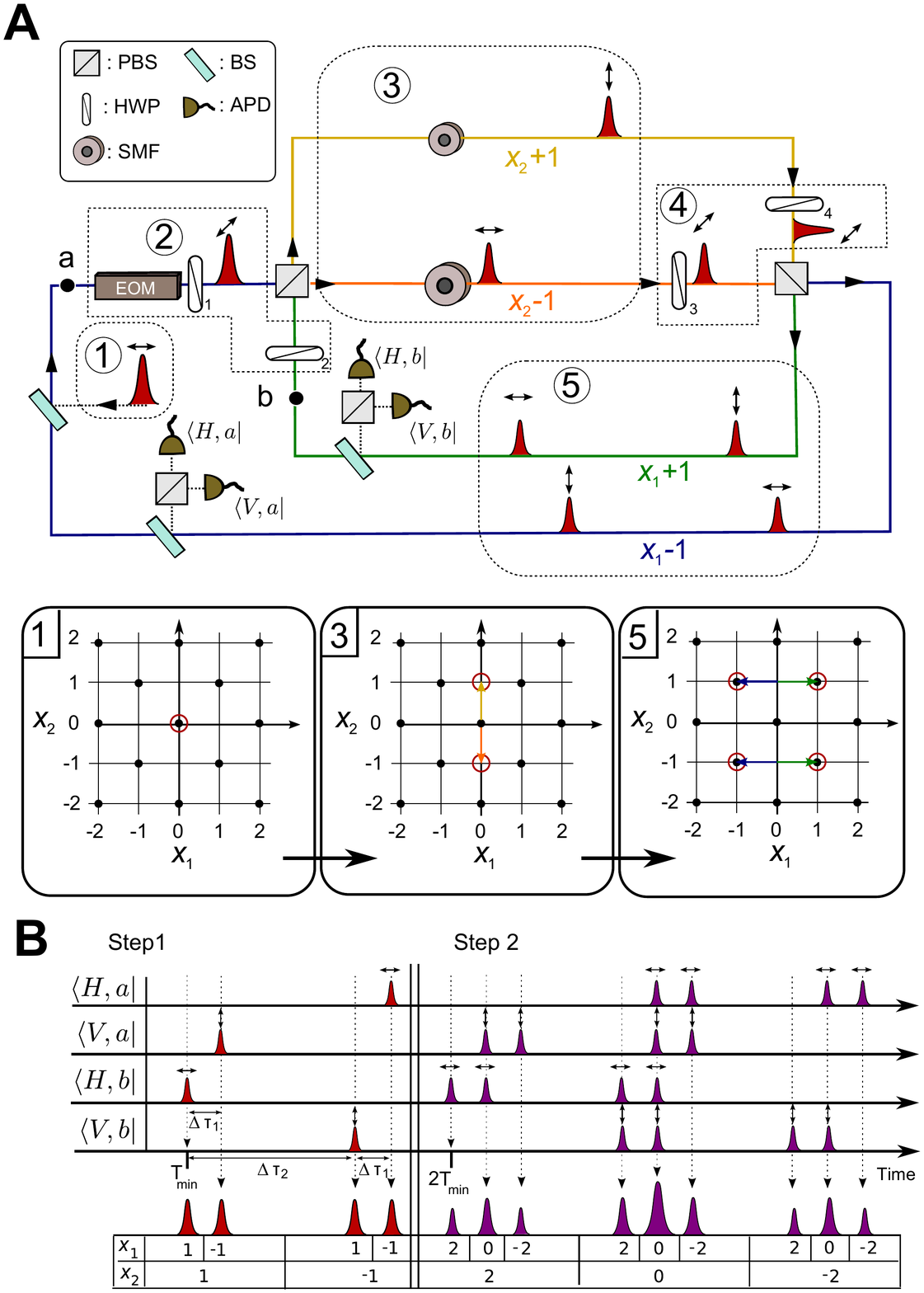}
\end{center}
\caption{ (\textbf{A}) Schematic setup including five time stages during the first step of a Hadamard walk. The laser pulse is initialized at (1) in horizontal polarization (arrow). After the coin operation  $\hat{C}_2$ (2) the vertical step (3) is implemented via two single-mode fibers (SMF) of different lengths. Subsequently  $\hat{C}_1$ (4) performs the coin operation for the horizontal direction, followed by the step operation implemented in free space (5). We compensate for different losses in arms a and b via different splitting ratios of the according BS. The corresponding spatial spread on a 2D lattice is shown for part (1), (3) and (5). (\textbf{B}) Mapping of the timing information of detected events onto spatial coordinates. The minimal time for a round trip $T_{min}$ is 676ns. For more details see text.}
\end{figure*}

Fig.~A.1 shows the first step of a quantum walk in our system and the corresponding mapping of the temporal spread on the 2D lattice.
The propagation of the initial pulse representing the walker through the setup is split into five parts:  

\begin{enumerate}
\renewcommand{\labelenumi}{(\theenumi)}
\item We initialize the photonic input state in the coin state $\ket{H,a}$ with the starting time corresponding to position $\ket{x_1,x_2} = \ket{0,0}$. 
\item After launching the pulse into the setup we perform the coin operation $\hat{C}_2 =\hat{C}_{\textrm{HWP}_2} \hat{C}_{\textrm{HWP}_1} \hat{C}_{\textrm{EOM}}$, as defined in Eq.~A2.
\item Split by a polarizing beam splitter (PBS) the pulse travels through two fibers of different lengths. The resulting time separation $\Delta\tau_2$ can be interpreted as a step in vertical ($x_2$) direction. We hereby define the transformation $x_2 \rightarrow x_2 -1$ for pulses passing the longer fiber, while the pulses in the shorter arm acquire the transformation $x_2 \rightarrow x_2+1$. Additional  retardation plates hereby compensate for unwanted polarization effects in the fibers.
\item Subsequently $\textrm{HWP}_3$ and $\textrm{HWP}_4$ implement the coin operation $\hat{C}_1 = \hat{C}_{\textrm{HWP}_3} \hat{C}_{\textrm{HWP}_4}$ (Eq.~A2),  determining the coin state for the horizontal ($x_1$) direction. 
\item The  step operator in the horizontal ($x_1$) direction is performed with a second PBS and a temporal shift $\Delta\tau_1$, obtained by traveling in two free space paths of different lengths.
\end{enumerate}
As a result one step in our quantum walk setup combines both a shift in vertical and horizontal direction, restricting the translation on a 2D grid to the diagonal neighbors (Fig.~1B). Starting in the origin, this leads to a spread over maximally $(n+1)^2$ positions after $n$ steps.  

After each step a time resolving measurement gives information about the location of the photon. The detection times for the first and second step of a Hadamard walk are shown in Fig.~A.1B. At the first detection the photons can arrive at four different times, which is either the minimal time $T_{\textrm{min}}$ (corresponding to position state $\ket{1,1}$),  $T_{\textrm{min}}+ \Delta\tau_1$  stating position $\ket{1,-1}$,  $T_{\textrm{min}}+ \Delta\tau_2$ ($\ket{-1,1}$) or  $T_{\textrm{min}}+ \Delta\tau_1+ \Delta\tau_2 $ ($\ket{-1,-1}$). At the second step on the other hand, multiple pulses can arrive simultaneously at the detectors. Independent of the coin state and, hence, spatial and polarization modes of the photons, all coinciding detections correspond to the same position state $\ket{x_1,x_2}$. In the following step wave packets in the same time bin and spatial mode interfere at $\textrm{HWP}_{1,2}$ ($\hat{C}_2$), while wave packets in different spatial modes can interfere at $\textrm{HWP}_{3,4}$ ($\hat{C}_1$).           

\section{Appendix 3: Limits and imperfections}\label{App. 3}
During the time evolution, the area of the grid covered by the quantum walker grows quadratically with the number of steps. 
The use of the time-multiplexing technique guarantees that the number of elements stays constant independent of the size of the simulated grid. In the experimental implementation it is only the lengths of the optical paths that needs to be adjusted to the maximum number of steps that are to be realized. In addition, the performance of our time-multiplexed setup is limited only by imperfections of the optical components resulting in errors, decoherence and losses. In the following we want to discuss each point individually. 

A sharp limitation for the maximal step number is given by the design of the experiment. If the minimal time for one round trip $T_{\textrm{min}}$  is shorter than the temporal expansion of all positions in a single step, temporal overlaps between different steps can occur. We choose our experimental parameters $T_\textrm{min}$ and $\Delta\tau_{1,2}$ without EOM such that $\Delta\tau_2 >13 \cdot\Delta\tau_1$ and $T_{\textrm{min}}>13 \cdot\Delta\tau_2$ to prevent temporal overlaps of different positions for the first 12 steps. An occurring additional delay induced by the modulator changed the conditions without inducing unwanted temporal overlaps. However, by simply changing the fiber lengths or the path differences the step number can easily be increased. 

The most significant source of systematic errors in the setup is the EOM. Due to the architecture of the modulator the applied phase is not only affecting the horizontal polarization component, as shown in Eq.~A2, but also the vertical component with a factor 1/3.5. This decreases the achievable similarities for quantum walks where controlled-Z coin operations are used. Additionally, the wave front of pulses passing the modulator are distorted differently for both polarizations, which influences the occurring interferences. Both effects are included in the theory presented in Fig.~3B. A replacement of the EOM with an optimized modulator would improve the achievable similarity at higher number of steps.

Decoherence in the time-multiplexed setup can occur if mechanical vibrations of the optical elements influence the interference properties. Typically mirrors vibrate with a frequency below 500Hz, corresponding to a time scale of 2ms. The duration of twelve steps  in the current setup is less than $10\mu s$, a factor of 200 faster compared to mirror vibrations. This suggests that decoherence effects will not influence the time-multiplexed quantum walk up to at least 100 steps.

At the present stage the main factor limiting the scalability is given by the losses per step. These are induced by the probabilistic detection method and losses at optical elements. To counter the effect of losses one can either start with an increased intensity or use optical amplifiers, as shown in (\textit{32}). While the first approach requires an active protection of the single-photon detectors, the second prohibits the use of the experiment with single-photon sources. A third method to reduce the losses is a change from probabilistic to a deterministic coupling mechanism with additional polarization modulators. This technique combined with a change-over to a low-loss wavelength regime (1550nm), makes the setup interesting for single photon input states. Using one of the described methods to circumvent the losses can increase the number of steps significantly.

\section{Appendix 4: Entanglement}\label{App. 4}
To quantify the two-particle entanglement simulated in the system we assumed that the quantum walk evolution results in a pure state $\ket{\psi_n}=\sum\limits_{x_1,x_2,c_1,c_2} a_{x_1,x_2,c_1,c_2}\ket{x_1,x_2,c_1,c_2}$ after $n$ steps, with the complex parameters  $a_{x_1,x_2,c_1,c_2} \in \mathbb{C}$. 
The assumption is based on the fact that the system does not show signs of decoherence for any of the coin operators, confirmed by the high values of the measured similarities. 

The von Neumann entropy $E$, which quantifies the entanglement (\textit{33}) is given by 
\begin{equation*}
E(\rho_1)=-\sum\limits_{i}\lambda_i \textrm{log}_2 \lambda_i,
\tag{A3}
\end{equation*} 
with the eigenvalues $\lambda_i$ of the reduced density matrix $\rho_1 = \textrm{Tr}_2(\rho) = \textrm{Tr}_2(\ket{\psi_n}\bra{\psi_n})$, given by the trace over one subsystem. For details see (\textit{23}).

By individually measuring the arrival probability in each coin and position state we obtain the absolute squared of the parameters $|a_{x_1,x_2,c_1,c_2}|^2$, hence no direct extraction of the phase information is possible. However, we can obtain information about the relative phases between position and coin states from the final interference pattern. As a result we can reconstruct the phase distribution with the help of the theoretical model up to three undetermined relative phases between the four different coin states and a global phase factor. By choosing the phases inducing the minimal entropy, we are able to give a lower bound for the simulated entanglement in the experiment.
\end{appendix}


\begin{thebibliography}{10}

\bibitem{Aharonov93}
Y.~Aharonov, L.~Davidovich, N.~Zagury, {\it Phys. Rev. A\/} {\bf 48}, 1687
  (1993).

\bibitem{Aharonov01}
D.~Aharonov, A.~Ambainis, J.~Kempe, U.~Vazirani, in {\it Proceedings of the
  Thirty-Third Annual ACM Symposium on Theory of Computing\/}, STOC '01, Hersonissos, Greece, 6 to 8 July 2001  (ACM,
  New York, NY, USA, 2001), pp. 50--59.

\bibitem{Kempe2003}
J.~Kempe, {\it Contemporary Physics\/} {\bf 44}, 307 (2003).

\bibitem{Konno2004}
N.~Inui, Y.~Konishi, N.~Konno, {\it Phys. Rev. A\/} {\bf 69}, 052323 (2004).

\bibitem{Kitagawa2010}
T.~Kitagawa, M.~S. Rudner, E.~Berg, E.~Demler, {\it Phys. Rev. A\/} {\bf 82},
  033429 (2010).

\bibitem{Mohseni2008}
M.~Mohseni, P.~Rebentrost, S.~Lloyd, A.~{Aspuru-Guzik}, {\it J. Chem. Phys.\/}
  {\bf 129}, 174106 (2008).

\bibitem{Plenio2008}
M.~B. Plenio, S.~F. Huelga, {\it N. J. Phys.\/} {\bf 10}, 113019 (2008).

\bibitem{Werner11}
A.~Ahlbrecht, {\it et~al.\/}, {\it arxiv: quant-ph/1105.1051\/}  (2011).

\bibitem{bouwmeester1999}
D.~Bouwmeester, I.~Marzoli, G.~P. Karman, W.~Schleich, J.~P. Woerdman, {\it
  Phys. Rev. A\/} {\bf 61}, 013410 (1999).

\bibitem{Schreiber10}
A.~Schreiber, {\it et~al.\/}, {\it Phys. Rev. Lett.\/} {\bf 104}, 050502
  (2010).

\bibitem{Broome10}
M.~A. Broome, {\it et~al.\/}, {\it Phys. Rev. Lett.\/} {\bf 104}, 153602
  (2010).

\bibitem{Schreiber11}
A.~Schreiber, {\it et~al.\/}, {\it Phys. Rev. Lett.\/} {\bf 106}, 180403
  (2011).

\bibitem{schaetz09}
H.~Schmitz, {\it et~al.\/}, {\it Phys. Rev. Lett.\/} {\bf 103}, 090504 (2009).

\bibitem{karski09}
M.~Karski, {\it et~al.\/}, {\it Science\/} {\bf 325}, 174 (2009).

\bibitem{zaehringer10}
F.~Z\"{a}hringer, {\it et~al.\/}, {\it Phys. Rev. Lett.\/} {\bf 104}, 100503
  (2010).

\bibitem{Sanders}
T.~D. Mackay, S.~D. Bartlett, L.~T. Stephenson, B.~C. Sanders, {\it J. Phys.
  A\/} {\bf 35}, 2745 (2002).

\bibitem{Childs09}
A.~M. Childs, {\it Phys. Rev. Lett.\/} {\bf 102}, 180501 (2009).

\bibitem{Lovett2010}
N.~B. Lovett, S.~Cooper, M.~Everitt, M.~Trevers, V.~Kendon, {\it Phys. Rev.
  A\/} {\bf 81}, 042330 (2010).

\bibitem{childs2003}
A.~M. Childs, {\it et~al.\/}, in {\it Proceedings of the thirty-fifth {ACM}
  symposium on Theory of computing} - {STOC} '03 (San Diego, {CA}, {USA},
  2003), p.~59.

\bibitem{SKW}
N.~Shenvi, J.~Kempe, K.~B. Whaley, {\it Phys. Rev. A\/} {\bf 67}, 052307
  (2003).

\bibitem{ambainis2004}
A.~Ambainis, J.~Kempe, A.~Rivosh, in {\it Proceedings of the Sixteenth ACM-SIAM Symposium on Discrete Algorithms}, {SODA '05}, Vancouver, Canada, 23 to 25 January 2005 (Society for Industrial and Applied Mathematics, Philadelphia, 2005), p.
  1099-1108.

\bibitem{Hillery2010}
M.~Hillery, D.~Reitzner, V.~Bu\v{z}ek, {\it Phys. Rev. A\/} {\bf 81}, 062324
  (2010).

\bibitem{Peruzzo10}
A.~Peruzzo, {\it et~al.\/}, {\it Science\/} {\bf 329}, 1500 (2010).

\bibitem{OBrien2011}
J.~C.~F. Matthews, {\it et~al.\/}, {\it arxiv: quant-ph/1106.1166\/}  (2011).

\bibitem{Osellame_2011}
L.~Sansoni, {\it et~al.\/}, {\it Phys. Rev. Lett.\/} {\bf 108}, 010502 (2012).

\bibitem{Wang11}
S.~D. Berry, J.~B. Wang, {\it Phys. Rev. A\/} {\bf 83}, 042317 (2011).

\bibitem{Spreeuw}
R.~J.~C. Spreeuw, {\it Found. Phys.\/} {\bf 28}, 361 (1998).

\bibitem{Kwiat1998}
N.~J. Cerf, C.~Adami, P.~G. Kwiat, {\it Phys. Rev. A\/} {\bf 57}, R1477 (1998).

\bibitem{Spreeuw2002}
N.~Bhattacharya, H.~B. van Linden van~den Heuvell, R.~J.~C. Spreeuw, {\it Phys.
  Rev. Lett.\/} {\bf 88}, 137901 (2002).

\bibitem{Roldan2005}
E.~Roldan, J.~C. Soriano, {\it J. Mod. Opt.\/} {\bf 52}, 2649 (2005).

\bibitem{Lahini2011}
Y.~Lahini, {\it et~al.\/}, {\it arxiv: quant-ph/1105.2273\/}  (2011).

\bibitem{hamilton11}
C.~S. Hamilton, A.~G\'{a}bris, I.~Jex, S.~M. Barnett, {\it N. J. Phys.\/} {\bf
  13}, 013015 (2011).

\bibitem{Reck1994}
M.~Reck, A.~Zeilinger, H.~J. Bernstein, P.~Bertani, {\it Phys. Rev. Lett.\/}
  {\bf 73}, 58 (1994).

\bibitem{Peschel11}
A.~Regensburger, {\it et~al.\/}, {\it Phys. Rev. Lett.\/} {\bf 107}, 233902
  (2011).

\bibitem{Barnett1989}
S.~M. Barnett, S.~J.~D. Phoenix, {\it Phys. Rev. A\/} {\bf 40}, 2404 (1989).

\end{thebibliography}
\end{document}